\documentclass[doublecol]{epl2}

\title{Signature of proximity induced $p_x+ip_y$ triplet pairing
in the doped topological insulator $\rm Bi_2Se_3$
by the s-wave superconductor NbN}
\shorttitle{Title} 

\author{Gad Koren\inst{1}, Tal Kirzhner\inst{1}, Yoav Kalcheim\inst{2} \and Oded Millo\inst{2}}

\institute{
  \inst{1} Physics Department, Technion - Israel Institute of Technology, Haifa 32000, Israel\\
  \inst{2} Racah Institute of Physics and the Hebrew University Center for Nanoscience
 and Nanotechnology, The Hebrew University of Jerusalem, Jerusalem 91904, Israel
}

\abstract{
 In the search for Majorana fermions in proximity induced topological superconducting junctions, we happened to find a signature of same-spin triplet superconductivity which appears to dominate these elusive elementary excitations. Thin film junctions and bilayers of the doped topological insulator $\rm Bi_2Se_3$ and the s-wave superconductor NbN exhibit conductance spectra with coexisting prominent zero bias and coherence peaks. Various tunneling models with different pair potentials have failed to fit our data, except for the triplet $p_x+ip_y$ pair potential, which breaks time reversal symmetry, that yielded reasonably good fits. This provides supporting evidence for proximity induced triplet superconductivity in the $\rm Bi_2Se_3$ layer near the interface with the NbN film.}

\begin{document}

\maketitle

\section{Introduction}
Topological superconductors (TSC) are interesting since they are predicted to support Majorana fermions (MF) which are protected against disorder and decoherence, and are therefore potential candidates to play an important role in future quantum computers \cite{KaneRMP,Kitaev,Oreg}. This type of superconductors can be realized by either doping of a topological insulator (TI) \cite{Hor,Ando,Tal}, or by putting a TI or a semiconductor with strong spin orbit coupling in good contact with a conventional superconductor \cite{LiLu,Koren1,Koren2}. In the former case, an intrinsic bulk superconductivity occurs, while in the second case, superconductivity is induced by the proximity effect. Recently, supporting experimental evidence for the existence of TSC and MF bound states was found in superconducting nano-wires of the proximity induced hybrids of InSb/NbTiN and InAs/Al \cite{Kouwenhoven,Heiblum}. Attempts of realistic modeling of these experimental results however, failed to reproduce the exact shape of the observed zero bias conductance peak (ZBCP) and its behavior under field, thus pointing to possible coexisting phenomena besides MF in a TSC \cite{Rainis}. In junctions of other TSC materials such as $\rm Cu_xBi_2Se_3$, $\rm Bi_2Se_3-Sn$, $\rm Bi_2Te_3-Bi$ and $\rm Bi_2Se_3-NbN$  with normal metals, generally robust ZBCP were observed, but conclusive evidence that these are due to MF in a TSC rather than Andreev bound states (ABS) in an unconventional superconductor, is still lacking \cite{Ando,Tal,LiLu,Koren1,Koren2}.\\

In the present study we extended our previous investigation of $\rm Au-Bi_2Se_3-NbN$ junctions of large overlap area ($\sim 100\times 150\, \mu m^2$), to smaller ramp-type junctions of a much smaller area ($5\times 0.5\, \mu m^2$), and to even smaller area junctions of a few nm diameter, resulting from controlled crashing of an STM tip into $\rm Bi_2Se_3-NbN$ bilayers. This was done in order to find out if the observed spectra are robust and independent of the junction area, and how the different crystallographic orientations (in particular in the ramp-junctions) affect the observed ZBCP and coherence peaks. The observed ZBCP disappeared above a proximity induced transition to superconductivity at $T_c$(junction)$\sim$8 K, while the coherence peaks survived up to $T_c$ of the NbN electrode at $\sim$10 K. The conductance spectra at low temperatures, which showed both ZBCP and coherence peaks, could be fitted very well with a minimal number of fitting parameters only when using the triplet $p_x+ip_y$ pair potential, which breaks time reversal symmetry. In a recent study by Nagai et al. \cite{Nagai},  a clear correspondence between topological superconductivity and spin triplet superconductivity was demonstrated. This provides a theoretical basis for using an existing triplet model \cite{Yamashiro}, to fit our present conductance spectra. We note that indications for spin-triplet superconductivity were found recently also in single crystals of the doped topological insulator $\rm Cu_xBi_2Se_3$ under high pressure and in point contact measurements \cite{Bay,Chen}.\\

\section{Preparation of the thin film junctions}
Preparation and characterization of the NbN, Au and $\rm Bi_2Se_3$ thin films by laser ablation deposition was described in our previous study \cite{Koren2}. All films and bilayers were deposited on (100) $\rm SrTiO_3$ (STO) wafers of $10\times10\, mm^2$ area. Here we shall only elaborate on the preparation of the fully photo-lithographically patterned ramp-type junctions, whose schematic cross-section is shown in the inset of fig.~\ref{fig.5}, and are also described in \cite{Nesher}. First, a bilayer of the base electrode consisting of an insulating and smooth STO layer of 50 nm thickness was deposited under vacuum at 150 $^0$C on a 70 nm thick NbN layer. This bilayer was then patterned in the form of fig. 2(c) of Ref. \cite{Koren2} by Ar ion milling on half the wafer, and left in ambient air for a day to create the 1-2 nm thick tunneling barrier of the native oxide layer ("nox" - marked by the inclined red line in the inset of fig.~\ref{fig.5}) \cite{Darlinski}. Then the cover electrode of 100 nm Au on 70 nm $\rm Bi_2Se_3$ was deposited and patterned to produce ten ramp junctions on the wafer, each with about $5\times 0.5\, \mu m^2$ junction area. Note that no current could flow  via the thick STO layer in the $5\times 2\, \mu m^2$ overlap area of the cover and base electrodes in the direction normal to the wafer. Conductance spectra were measured by the four-probe technique in the same cooling run on all the ten junctions, using a $4\times 10$ array of spring loaded gold coated tips pressed on the 40 gold contact pads on the wafer. \\

\section{Experimental results and fits of the data}

\begin{figure} \hspace{-3mm}
\onefigure[height=6.5cm,width=8.5cm]{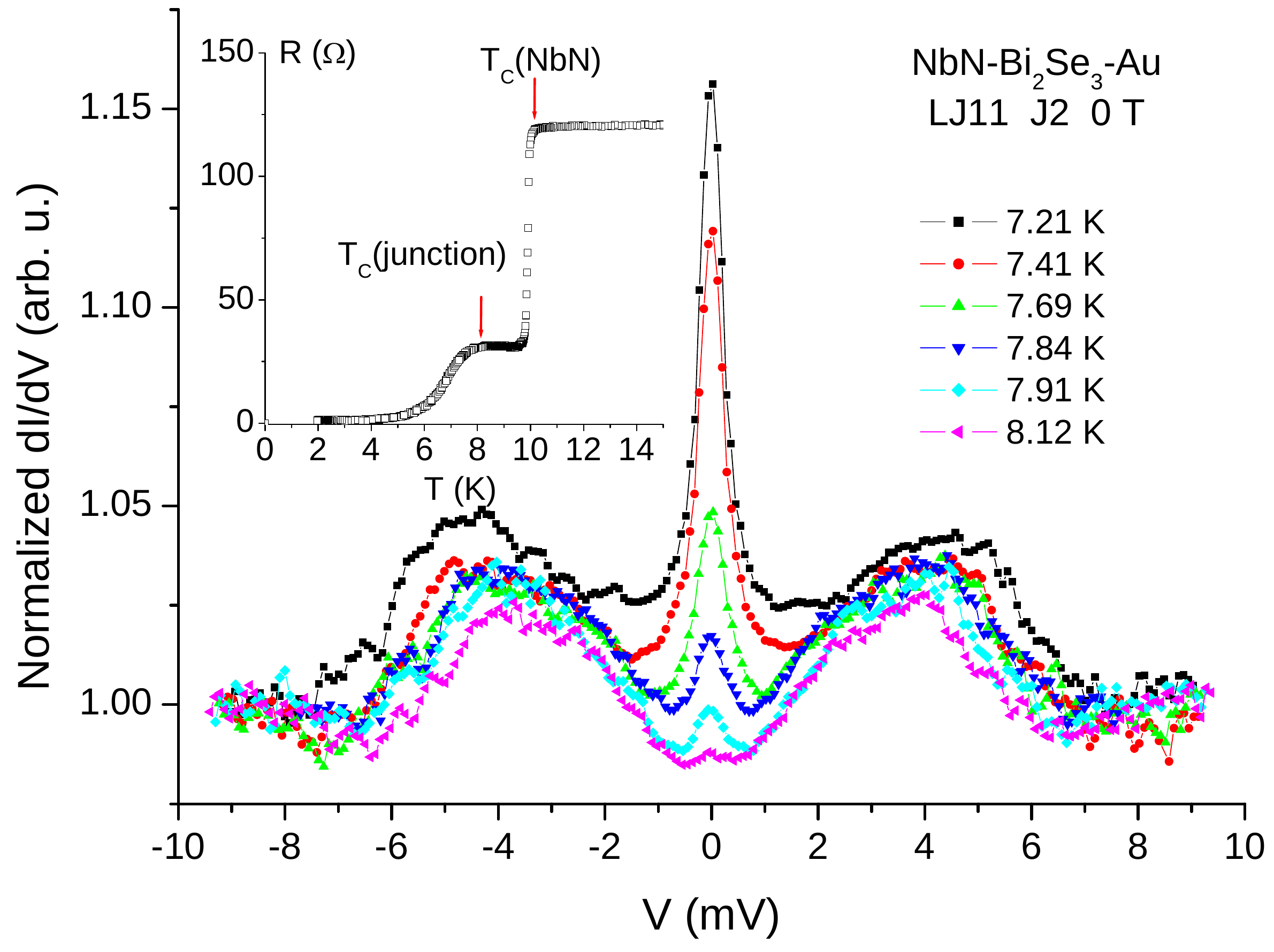}
\vspace{-0mm} \caption{ (Color online) Normalized conductance spectra of a large junction with $100\times 30\,\mu m^2$ overlap area at various temperatures.  The overlapping layers of this junction (as in fig. 2(a) of Ref. \cite{Koren2}) are: 100 nm Au on 20 nm $\rm Bi_2Se_3$ on 1-2 nm native oxide barrier on 70 nm NbN. The inset shows the resistance versus temperature of this junction, with the transition of the NbN electrode at $T_c(\rm NbN)$ and the proximity induced transition of the junction at $T_c(\rm junction)$.       }
\label{fig.1}
\end{figure}

First, we demonstrate that ZBCPs in the junctions were observed only below the transition temperature of the proximity induced superconducting layer of the $\rm Bi_2Se_3$ in the interface region with the NbN film. Fig.~\ref{fig.1} shows normalized conductance spectra at different temperatures of a large junction with overlap area of $100\times 30\,\mu m^2$, together with the resistance versus temperature of this junction in the inset. Below the transition to superconductivity of the NbN electrode at $T_c(\rm NbN)\sim$10 K, one can see a clear second transition at about 8 K which is due to the proximity effect in the $\rm Bi_2Se_3$ layer near the interface with the NbN layer. This second transition therefore originates in the junction and is denoted by $T_c(\rm junction)$.  At low temperature the ZBCP is robust (not shown). With increasing temperature it decreases in height but persists up to $T_c(\rm junction)\sim$8 K, above which it disappears and only the weak coherence peaks remain up to $T_c$ of the NbN electrode. A clear correlation is therefore shown between the presence of ZBCPs in the main panel and the proximity induced superconductivity in the junction.\\

\begin{figure} \hspace{-3mm}
\onefigure[height=6.5cm,width=8.5cm]{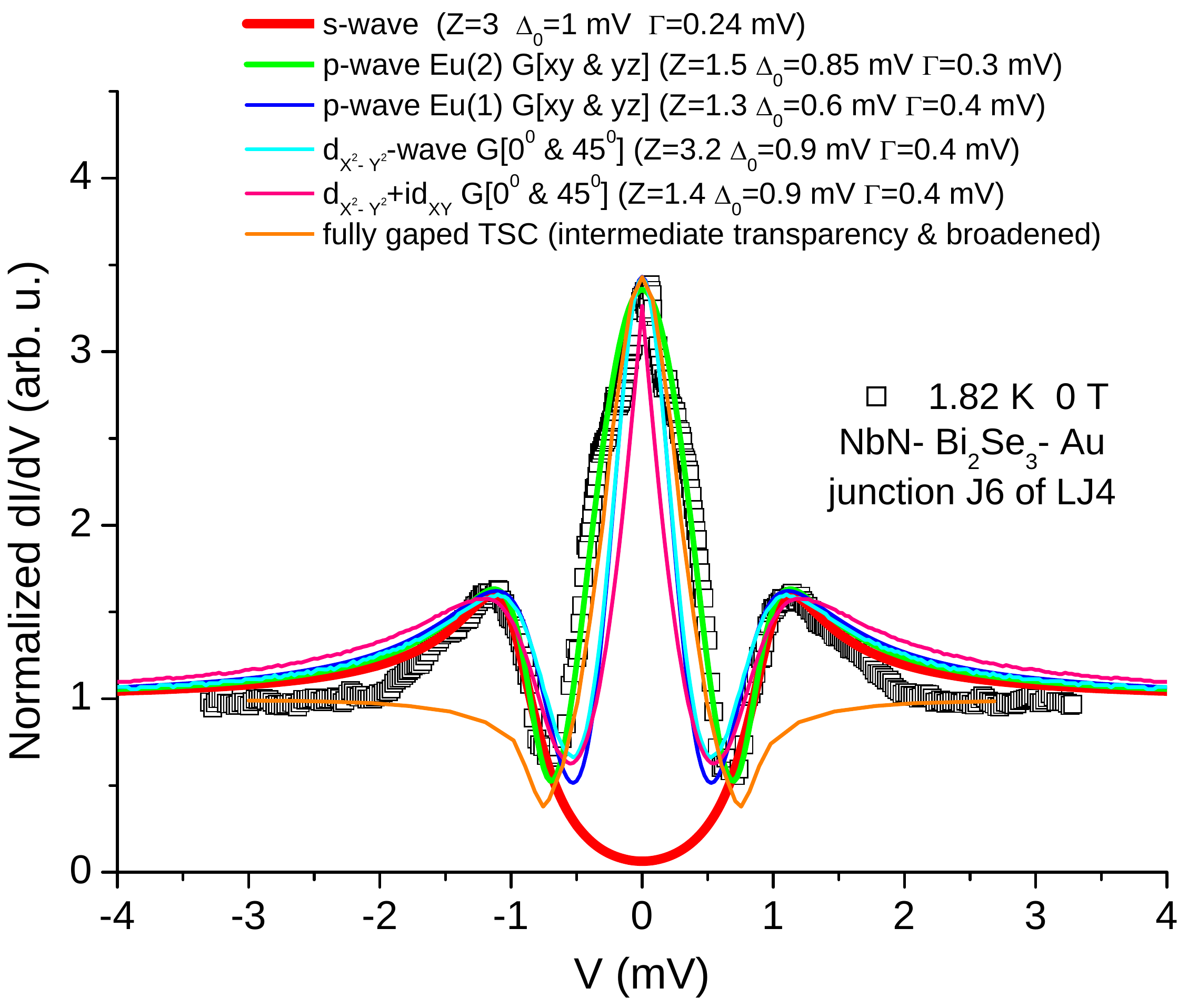}
\vspace{-0mm} \caption{ (Color online) Normalized conductance spectra of a large junction of a previous study of our group (fig. 5(d) of Ref. \cite{Koren2}), together with fits to different theoretical models and pair potentials.  }
\label{fig.2}
\end{figure}

Next, we use a conductance spectrum  with a robust ZBCP and coherence peaks from our previous study (fig. 5(d) of Ref. \cite{Koren2}) as a test case for finding which pair potential (PP) fits this data best with a minimum number of fitting parameters. Fitting was done using the Blonder, Tinkham and Klapwijk (BTK) model \cite{BTK}, with modifications of it to other PP symmetries as needed \cite{Tanaka,Yamashiro,Yamakage}. The standard parameters of such fits, as applied also here, are the barrier strength Z, the pairing amplitude or energy gap $\Delta_0$, and the finite lifetime broadening $\Gamma$. Various pair potentials were used, and the results of the fits are shown in fig.~\ref{fig.2} together with the original experimental data. Singlet s-wave pair potential could fit the coherence peaks but not the ZBCP, while a fully gapped TSC, as well as a TSC with point nodes, could barely fit the ZBCP and the two dips beside it, but not the coherence peaks \cite{Yamakage}. We note that since there is no direct contact between the NbN electrode and the gold overlayer, the observed coherence peaks in fig.~\ref{fig.2} could not originate in any NbN-Au junction. It is also noted that the conductance spectra predicted by Yamakage et al. \cite{Yamakage}, who used various PP as proposed by Fu and Berg \cite{FuBerg}, had never shown coherence peaks besides the ZBCP. Moreover, if we assume the existence of two bands of the surface and bulk states in the superconducting $\rm Bi_2Se_3$ near the interface, with two different $\Delta_s$ and $\Delta_b$ gaps, we could fit the data of fig.~\ref{fig.2} quite well using the s-wave PP. This however necessitates very different barrier strengths (Z values) for the two bands which is highly unlikely, but it can not be ruled out completely as the scattering rates for these bands could be different. Nevertheless, in order to keep within the BTK model with a minimum number of fitting parameters, we tried other PPs such as the singlet d-wave $d_{x^2-y^2}$ and the mixed $d_{x^2-y^2}+id_{xy}$ \cite{Tanaka}, ignoring the hexagonal symmetry of $\rm Bi_2Se_3$, but compensating for it by taking weighted tunneling contributions from the node and antinode directions while keeping the same fitting parameters Z, $\Delta_0$ and $\Gamma$ for both orientations. This weighted sum of the calculated individual conductances along the anti-node $G(0^0)$ and node $G(45^0)$ directions amounts to $a[G(0^0)-1]+b[G(45^0)-1]+1$ where $a$ and $b$ are fit parameters. They represent the relative weight of the two orientations $b/a$, and the normalization factor to the experimental data $a$. Fig.~\ref{fig.2} shows that the fits improved, but not sufficiently to reproduce the whole width of the measured ZBCP. Ignoring the hexagonal symmetry of the  $\rm Bi_2Se_3$ layer on the cubic NbN film can be further justified by noting its mosaic structure as shown in fig. 12 of Ref. \cite{Koren2}, and also by noting that the for low orbital angular momentum and $k_z$ near zero, the tetragonal and hexagonal symmetries yield similar results \cite{Black}. Finally, we tried the odd parity triplet PP $p_x+ip_y$ which breaks time reversal symmetry, but shows consistency with some experimental results \cite{Bay,Chen}. Following Yamashiro, Tanaka and Kashiwaya \cite{Yamashiro}, we used $\Delta_{\uparrow\uparrow}=\Delta_0 sin\theta(cos\phi + sin\phi)$ and
$\Delta_{\uparrow\uparrow}=\Delta_0 sin\theta(cos\phi + isin\phi)$ with $\theta$ and $\phi$ being the polar and azimuthal angles, for the two (tetragonal) symmetries $\rm E_u(1)$ and $\rm E_u(2)$, respectively, while $\Delta_{\downarrow\downarrow}=\Delta_{\uparrow\downarrow}=\Delta_{\downarrow\uparrow}=0$. As before, we sum over weighted contributions to the conductance from two different interfaces (xy and yz, where xy represents the a-b plane of the $\rm Bi_2Se_3$ layers), and keep the same basic fitting parameters Z, $\Delta_0$ and $\Gamma$ for both interfaces.  The justification for taking the weighted contributions of the two interfaces is that both are present in our junctions, either via the roughness of the interfaces or due to the inclined geometry of the barrier in the ramp junctions. Fig.~\ref{fig.2} shows that while the best fit using the $E_u(1)$ PP fails to fit the ZBCP, the $E_u(2)$ PP fits the data quite well. All the features of the spectrum are captured satisfactorily and only small deviations are found which can be attributed to the tetragonal rather than hexagonal symmetry used in the model calculations. It is thus concluded that if this good fit using the triplet $p_x+ip_y$ wave PP has any physical significance rather than just being coincidental, it should also fit conductance spectra of other $\rm Bi_2Se_3$-NbN junctions of various types. In the following we show that this is actually the case in both ramp-type junctions where the combined contributions to the conductance from the xy and yz planes is evident, and point contact junctions which probe a highly transparent $\rm Bi_2Se_3-NbN$ interface. It is noted that in large junctions, it is hard to determine if the ZBCP arises from the interface roughness or originates in via-holes in the NbN layer, both of which add yz interfaces to the dominant xy ones \cite{Koren2}. We therefore chose to use smaller junction, such as the ramp junctions, with their well defined geometry and clear xy and yz interfaces.\\

\begin{figure} \hspace{-3mm}
\onefigure[height=6.5cm,width=8.5cm]{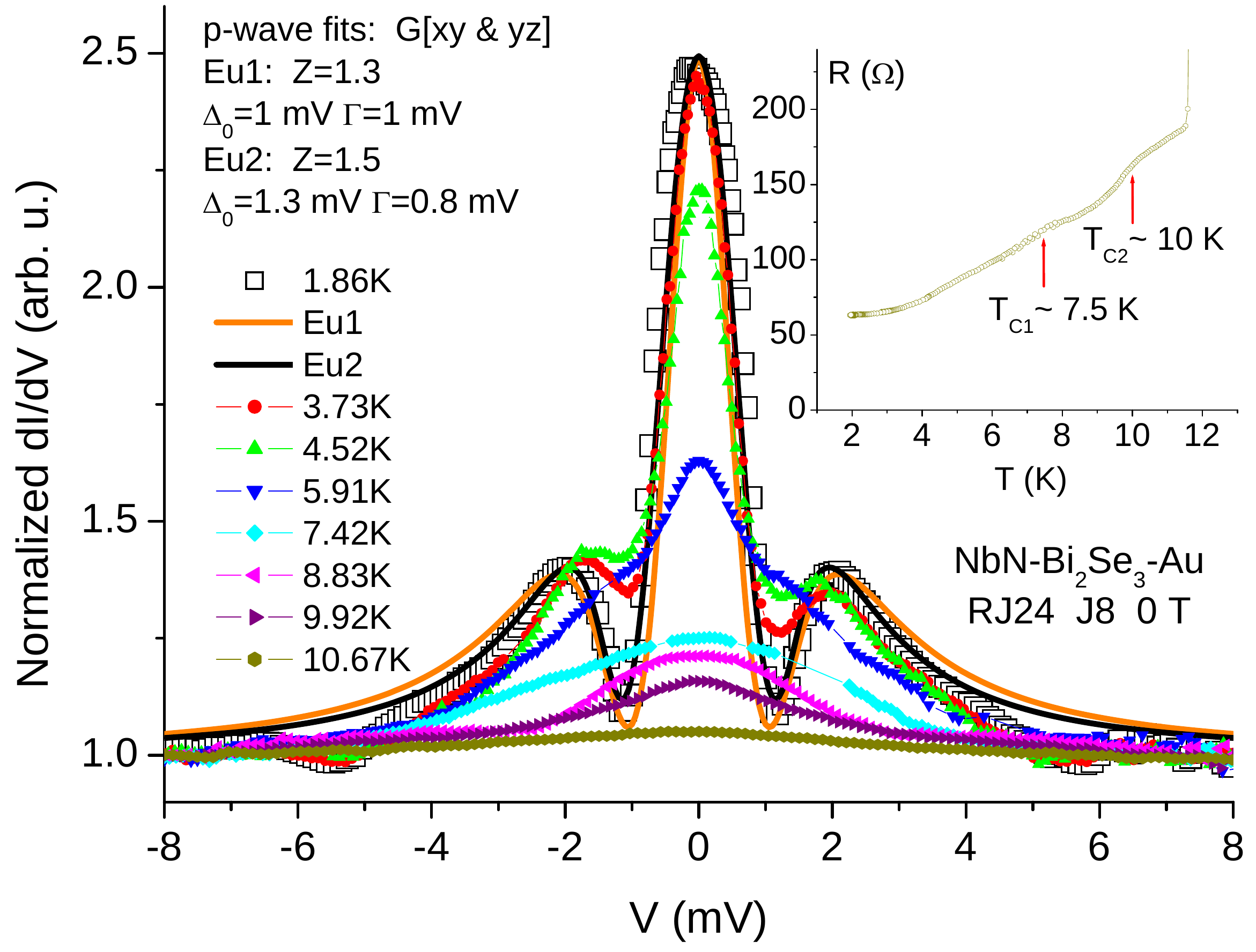}
\vspace{-0mm} \caption{ (Color online) Normalized conductance spectra of a small ramp type junction of $5\times 0.5\,\mu m^2$ area at various temperatures, together with two p-wave fits at 1.86 K. The inset shows the resistance versus temperature of this junction, where the two kinks coincide with the temperatures $T_{c1}$ and $T_{c2}$ at which the narrow and broad ZBCPs disappear, respectively.  }
\label{fig.3}
\end{figure}

\begin{figure} \hspace{-3mm}
\onefigure[height=6.5cm,width=8.5cm]{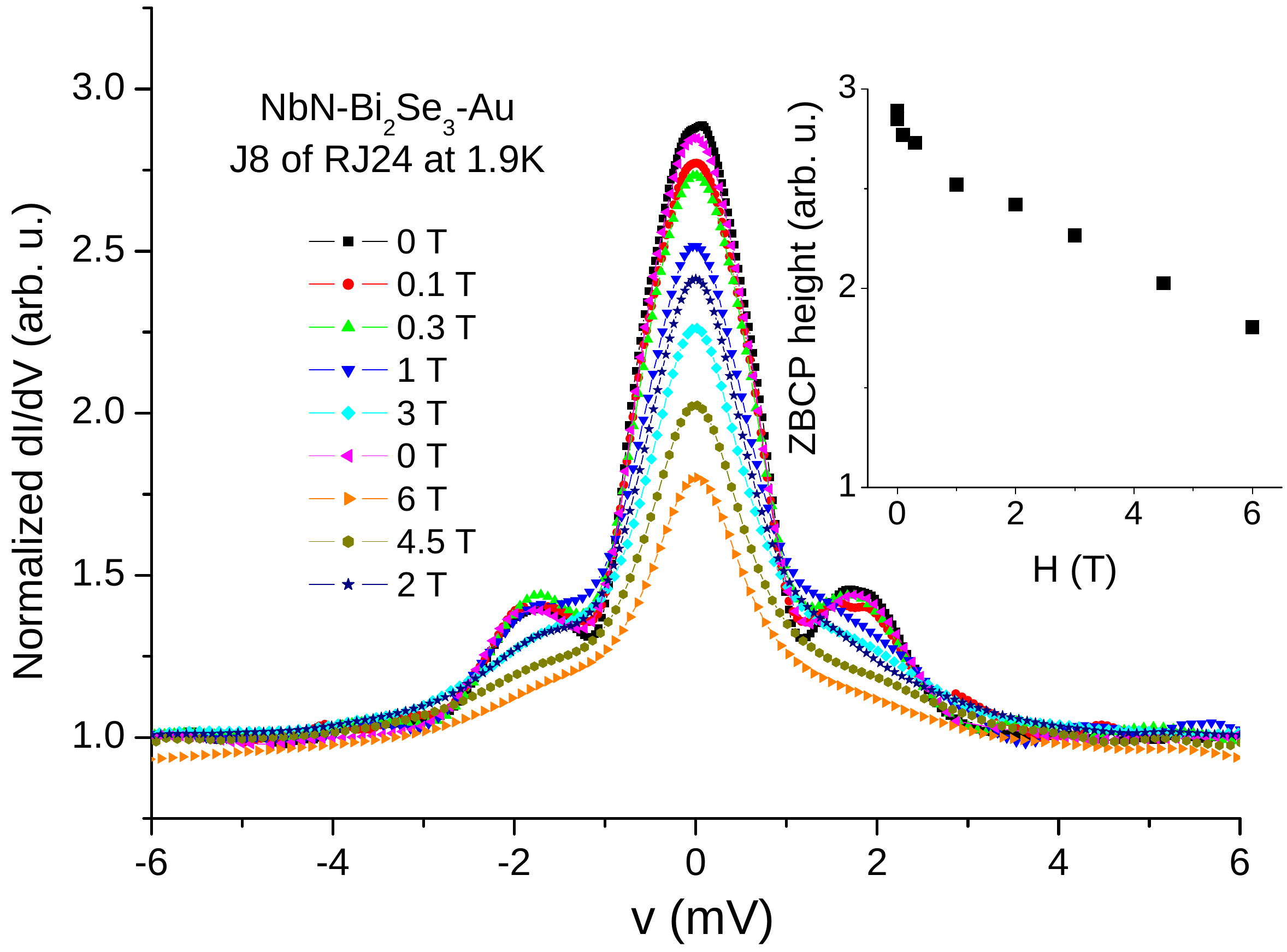}
\vspace{-0mm} \caption{ (Color online) Normalized conductance spectra of the same ramp type junction as in fig.~\ref{fig.3} at 1.9 K and under various magnetic fields. The inset shows the ZBCP heights versus field. }
\label{fig.4}
\end{figure}

\begin{figure} \hspace{-3mm}
\onefigure[height=6.5cm,width=8.5cm]{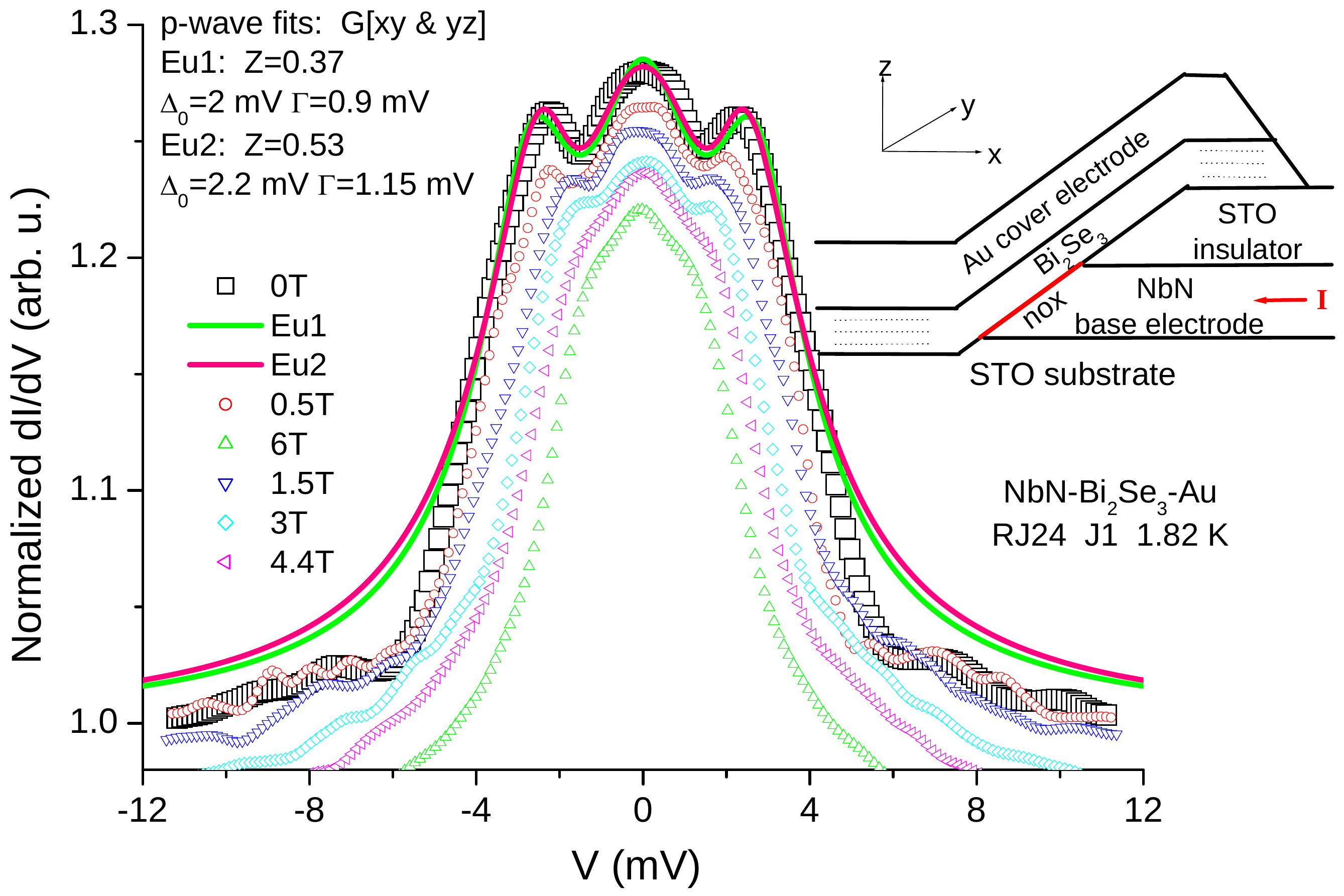}
\vspace{-0mm} \caption{ (Color online) Normalized conductance spectra at 1.82 K and different magnetic fields of another ramp junction on the same wafer as that of fig.~\ref{fig.3}. This junction is more transparent than that of fig.~\ref{fig.3}, as the corresponding Z parameters of the p-wave fits show. The inset depicts a schematic cross section of a typical ramp junction.  }
\label{fig.5}
\end{figure}

Figs.~\ref{fig.3}, ~\ref{fig.4} and ~\ref{fig.5} show normalized conductance spectra obtained on ramp junctions whose cross section is shown schematically in the inset of fig.~\ref{fig.5}. At low temperature and zero magnetic  field, figs.~\ref{fig.3} and ~\ref{fig.4} show tunneling-like spectra with robust ZBCP and coherence peaks similar to those of fig.~\ref{fig.2}, while fig.~\ref{fig.5} shows Andreev-like spectra, but still with clear ZBCP and coherence peaks. The low temperature spectra of these figures could be fitted reasonably well using the triplet $E_u(1)$ and $E_u(2)$ PP, with a slight preference to the $E_u(2)$ PP. Fig.~\ref{fig.3} also shows spectra at different temperatures that exhibit interesting correlation with the resistance versus temperature results presented by the inset of this figure where two kinks are found at $T_{c1}\sim$7.5 K and $T_{c2}\sim$10 K. The robust ZBCP of $\sim$1 mV width persists up to about $T_{c1}$, while above it and up to $T_{c2}$ only a broader ZBCP of $\sim$2 mV width remains. This is apparently due to the merging of the decaying ZBCP and coherence peaks with increasing temperature, but could also result from different surface and bulk contributions at the chemical potential to the conductance \cite{Tal}, as will be further discussed below. Figs.~\ref{fig.4} and ~\ref{fig.5} show that with increasing magnetic field at low temperature the spectra become narrower, decay in height, and smeared as the ZBCP and coherence peaks merge with one another. The decreasing conductance at high bias and high fields in fig.~\ref{fig.5} is due to flux flow. Fig.~\ref{fig.4} shows some magnetic hysteresis effects, which are apparently also responsible for the weaker dips as compared to those of the spectra in fig.~\ref{fig.3} where the data was obtained in a different cooling run without any application of magnetic fields. In all our junctions we had never seen a ZBCP increasing with field as in Refs. \cite{Kouwenhoven,Heiblum}. From the good fits of the data in fig.~\ref{fig.3} and ~\ref{fig.5} we conclude that also in small ramp junctions with different transparencies, the triplet $p_x+ip_y$ PP describes our results well. \\

\begin{figure} \hspace{-3mm}
\onefigure[height=6.5cm,width=8.5cm]{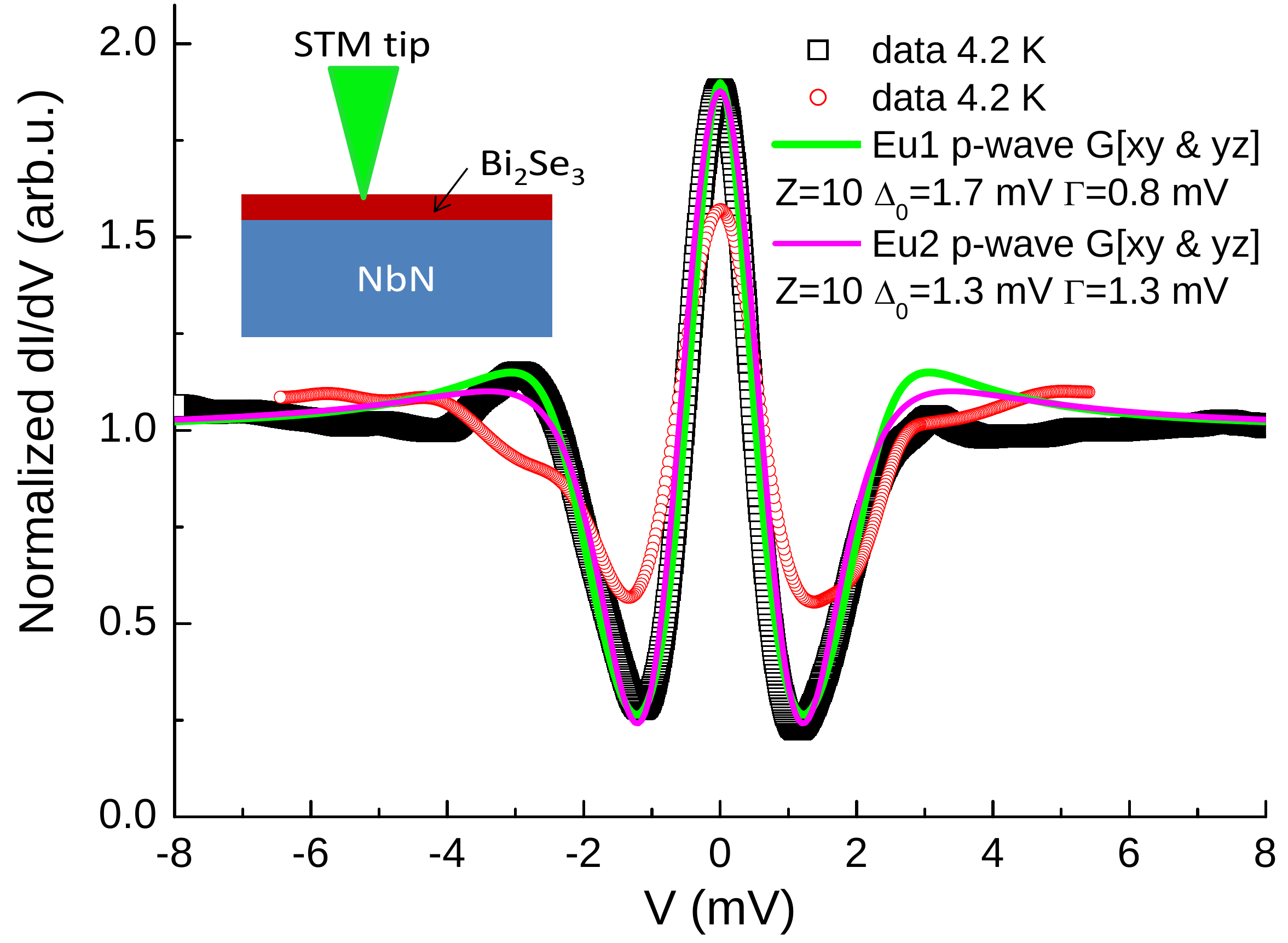}
\vspace{-0mm} \caption{ (Color online) Two normalized conductance spectra at 4.2 K of a point contact junction of a Pt/10\%Ir STM tip which was intentionally crashed into a bilayer of 10 nm $\rm Bi_2Se_3$ on top of 70 nm NbN, as seen in the inset. The curves show fits to the Eu(1) and Eu(2) p-wave model. }
\label{fig.6}
\end{figure}

Fig.~\ref{fig.6} depicts two normalized conductance spectra at 4.2 K of a point contact junction on an \textit{in-situ} prepared bilayer of a 10 nm thick $\rm Bi_2Se_3$ layer on top of a 70 nm NbN film (no native oxide at the $\rm Bi_2Se_3$-NbN interface). The wafer consisted of two halves, one half with the bilayer and the other half with a reference bare film of 70 nm NbN. In the STM mode (vacuum gap between tip and sample), the tunneling spectra measured on both halves did not reveal clear signatures of superconductivity, apparently due to a high Z oxide layer on the NbN, and the presence of a highly deteriorated top layer of the $\rm Bi_2Se_3$  which developed when the sample was exposed to ambient air. Only when the Pt/10\%Ir STM tip was intentionally and gradually crashed into the top of the $\rm Bi_2Se_3$ layer of the bilayer, forming a point contact of a few nm diameter, the spectra of fig.~\ref{fig.6} could be measured. Again, use of the triplet $p_x+ip_y$ PP fitted the data very well, with both the $E_u(1)$ and $E_u(2)$ PP and a Z=10 value, the highest used in the present study. In these fits there is a slight preference to the $E_u(1)$ PP as it fits better also the left hand side coherence peak and uses a much lower lifetime broadening $\Gamma$ value. Here again, we come to the same conclusion that the triplet $p_x+ip_y$ PP describes our data well, also in this very small point contact junction of a few nm diameter in close proximity to a highly transparent $\rm Bi_2Se_3-NbN$ interface.\\

\section{Discussion}

Fu and Kane had discussed the issue of a proximity effect between an s-wave superconductor and a topological insulator \cite{FuKane}. They predicted that the superconductor induces in the surface states of the TI  a superconducting PP which resembles that of a spinless $p_x+ip_y$ state. By a proper transformation of the base functions they showed that this PP could be made precisely spinless, and with it the Hamiltonian preserves time reversal symmetry. The presently used triplet $p_x+ip_y$ PP which fits our data nicely is clearly a spinfull PP which breaks time reversal symmetry. To reconcile these two seemingly contradictory results, we propose a scenario in which the odd-parity, interorbital spin-triplet $\hat{\Delta}_4$ PP of Fu and Berg \cite{FuBerg}, which breaks time reversal symmetry, is induced in the $\rm Bi_2Se_3$ near the interface together with the $\hat{\Delta}_2$ PP which preserves it and is supposed to exist in $\rm Cu_xBi_2Se_3$. $\hat{\Delta}_4$ is a spinfull $p_x+ip_y$ PP which is rotationally invariant around the z-axis under the lattice symmetry and is consistent with our present results. It is possible that the proximity induced $\hat{\Delta}_4$ PP originates in the bulk states at the chemical potential of our highly doped $\rm Bi_2Se_3$ films (see the ARPES data in Ref. \cite{Tal}), and if its contribution to the tunneling conductance at the interface is dominant, it might mask the surface contribution that could come from $\hat{\Delta}_2$. We note that in the proximity region of the interface both $\hat{\Delta}_4$ and $\hat{\Delta}_2$ can coexist, while in a TSC such as possibly $\rm Cu_xBi_2Se_3$, they could not since they belong to different irreducible representations. The observed ZBCPs are thus due to a spinfull triplet PP and apparently not to Majorana bound states as previously envisioned. We believe that if the bulk conductivity of our $\rm Bi_2Se_3$ films could be significantly reduced, observation of MF in our junctions could be realized. We also note that in most worldwide films and single crystals of $\rm Bi_2Se_3$ the bulk doping with Se vacancies is a major problem, which yields dominant metallic bulk conductance that masks the topological surface conductance. The latter of course, could yield MF bound states when in the superconducting phase. In the near future, we plan to measure low energy muon spin relaxation from bilayers of $\rm Bi_2Se_3$ on NbN in order to see if there is a net magnetization at zero field in the interface layer, originating in the same-spin triplet PP observed in the present study.\\

\section{Conclusions}

We found a signature of proximity induced triplet superconductivity by NbN in the doped topological insulator $\rm Bi_2Se_3$. Conductance spectra of various junctions with different transparencies that have robust ZBCPs and coherence peaks, could be fitted quite well using the chiral $p_x+ip_y$ pair potential. Thus, the observed ZBCPs of this odd parity, same-spin triplet pairs represent zero energy surface bound states which apparently do not originate from Majorana fermions.\\

\acknowledgments We acknowledge useful comments and discussions with Yukio Tanaka, Erez Berg and Assa Auerbach. This research was supported in part by the Israel Science Foundation, the joint German-Israeli DIP project, the US-Israel BSF, the Harry de Jur Chair in Applied Science (OM), and the Karl Stoll Chair in advanced materials at the Technion (GK).\\


\bibliography{apssamp}

\end{document}